\documentclass{pasj00}

\begin{document}
\SetRunningHead{Y. Takeda et al.}{Detection of Low-Level Activities in Solar-Analog Stars}
\Received{2012/05/28}
\Accepted{2012/06/25}

\title{Detection of Low-Level Activities in Solar-Analog Stars \\
from the Emission Strengths of Ca II 3934 Line
\thanks{Based on data collected at Subaru Telescope, which is operated
by the National Astronomical Observatory of Japan. }
}

%

\author{
Yoichi \textsc{Takeda},\altaffilmark{1}
Akito \textsc{Tajitsu},\altaffilmark{2}
Satoshi \textsc{Honda},\altaffilmark{3}
Satoshi \textsc{Kawanomoto},\altaffilmark{1}\\
Hiroyasu \textsc{Ando},\altaffilmark{1} and
Takashi \textsc{Sakurai}\altaffilmark{1}}
\altaffiltext{1}{National Astronomical Observatory, 2-21-1 Osawa, Mitaka, Tokyo 181-8588}
\email{takeda.yoichi@nao.ac.jp, kawanomoto.satoshi@nao.ac.jp, ando.hys@nao.ac.jp,\\ 
sakurai@solar.mtk.nao.ac.jp}
\altaffiltext{2}{Subaru Telescope, 650 North A'ohoku Place, Hilo, Hawaii 96720, U.S.A.}
\email{tajitsu@subaru.naoj.org}
\altaffiltext{3}{Kwasan Observatory, Graduate School of Science, Kyoto University,\\
17 Ohmine-cho Kita Kazan, Yamashina-ku, Kyoto 607-8471}
\email{honda@kwasan.kyoto-u.ac.jp}

\KeyWords{stars: activity --- stars: atmospheres --- 
stars: solar-type --- stars: rotation
}
\maketitle

\begin{abstract}
Activity studies of solar-type stars, especially with reference
to the status of our current Sun among them, have exposed the 
importance of (1) homogeneously selecting the sample stars and 
(2) reliably evaluating their activities down to a considerably
low level. Motivated by these requirements, we conducted an extensive 
study on the activities of 118 solar-analog stars (of sufficiently 
similar properties to each other) by measuring the emission
strength at the core of Ca~{\sc ii} 3933.663 line (K line) on
the high-dispersion spectrogram obtained by Subaru/HDS, where 
special attention was paid to correctly detecting the chromospheric
emission by removing the wing-fitted photospheric profile calculated from 
the classical solar model atmosphere. This enabled us to detect low-level 
activities down to $\log R' \sim -5.4$ ($R'$ is the ratio of the 
chromospheric core emission flux to the total bolometric flux), by which 
we could detect subtle activity differences which were indiscernible 
in previous studies. Regarding the Sun, we found $\log R'_{\odot} = -5.33$ 
near to the low end of the distribution, which means that it belongs to 
the distinctly low activity group among solar analogs. This excludes 
the once-suggested possibility for the high frequency of Maunder-minimum 
stars showing appreciably lower activities than the minimum-Sun. 
\end{abstract}

%


\section{Introduction}

It is of great interest for solar as well as stellar astrophysicists 
to compare the activity of our Sun to those of a number of other 
similar solar-type stars, since it may provide us with an opportunity 
to infer the trend of solar activity on a very long astronomical 
time scale (i.e., investigating the long-time behavior of a star 
may be replaced by studying many similar stars at a given time).

Baliunas and Jastrow (1990) argued based on the results of Mt. Wilson 
Observatory's HK survey project for 74 solar-type stars that the 
distribution of $S$-index
(nearly equivalent to 
$\propto \int_{\rm line}{F_{\lambda}}d\lambda /\int_{\rm cont}{F_{\lambda}}d\lambda$;
i.e., the ratio of integrated core-flux of Ca~{\sc ii} HK lines 
to the continuum flux; cf. Vaughan et al. 1978)
is bimodal, with about 1/3 showing appreciably smaller activities
than the Sun, which they interpreted as being in the ``Maunder-minimum''   
state of activity.  If this is true, it may mean that the spotless phase
of considerably low-activity such as occurred in late 17th century 
in our Sun (Eddy 1976) may not necessarily be an unusual phenomenon 
in the long run.

However, this result could not be confirmed by a similar analysis done by
Hall and Lockwood (2004), who reported based on many repeated observations
of Ca~{\sc ii} H and K lines for 57 Sun-like stars along with the Sun
at Lowell Observatory that such a bimodal distribution of $S$-index 
(as suggesting the existence of a considerable fraction of appreciably 
lower activity stars than the current Sun) is not observed; actually, 
even the $S$-values of 10 ``flat-activity'' stars turned out to be 
comparable with (or somewhat larger than) the typical solar-minimum value. 

Furthermore, Wright (2004) pointed out an important problem in 
Baliunas and Jastrow's (1990) sample selection. He concluded by examining
the absolute magnitudes of their sample based on Hipparcos parallaxes 
that many of those ``Maunder-minimum stars'' with considerably low $S$ 
indices are old stars evolved-off the main sequence, which suggests 
that their apparently low activity is nothing but due to the aging effect 
without any relevance to the cyclic or irregular change of activity
in solar-type dwarfs.   
Thus, fairly speaking, the original claim by Baliunas and Jastrow 
(1990) appears to be rather premature and difficult to be justified 
from the viewpoint of these recent studies.

Yet, the issue of clarifying the status of solar activity
among Sun-like stars does not seem to have been fully settled
and further investigations still remain to be done:\\
--- First, since understanding the activity of the Sun from a 
comprehensive perspective is in question, comparison samples 
should comprise stars as closer to the Sun as possible.
Admittedly, those authors surely paid attention to this point: 
Baliunas and Jastrow's (1990) HK project targets were in the 
$B-V$ range of 0.60--0.76, while Hall and Lockwood's (2004) 
Sun-like stars sample were chosen from stars of 
$0.58 \le B-V \le 0.72$, both narrowly encompassing the solar 
$(B-V)_{\odot}$ of 0.65 (Cox 2000). However, the homogeneity of 
these samples are not yet satisfactory. Could they be made up 
of further more Sun-like stars or solar analogs? \\
--- Second, it appears that the precision of detecting low-level 
activity has been insufficient. Although Mt. Wilson $S$ index 
reflects the core-emission strength (equivalent width) of 
Ca~{\sc ii} HK lines, it would not be a sensitive activity indicator 
any more, when the emission becomes weak, as it stabilizes
at a constant value determined by the photospheric absorption
profile. While an indicator for the pure-emission strength,
$R'_{\rm HK} (\equiv R_{\rm HK} - R_{\rm phot}$; where
$R_{\rm HK} \equiv F_{\rm HK}/F_{\rm bol}$ and
$R_{\rm phot} \equiv F_{\rm phot}/F_{\rm bol}$), has also been 
introduced to rectify this shortcoming and widely used,
the photospheric component $R_{\rm phot}$ is in most cases 
only roughly evaluated as a simple function of $B-V$ (e.g., 
Noyes et al. 1984) and thus its accuracy is rather questionable.
Wright (2004) also pointed out the importance of correctly
subtracting this component, in view of its possible dependence on
other stellar parameters (i.e., not only on $B-V$ or $T_{\rm eff}$, 
but also on $\log g$ and [Fe/H]).

These requirements motivated us to conduct a new investigation
on this subject, since we have been engaged these years with 
the extensive study of 118 Sun-like stars selected by the criteria
of $0.62 \ltsim B-V \ltsim 0.67$ and $4.5 \ltsim M_{V} \ltsim 5.1$
($M_{V,\odot} = 4.82$; Cox 2000), a quantitatively as well as 
qualitatively ideal sample of solar-analog stars. 
This project was originally started for the purpose of clarifying 
the behavior of Li abundances ($A$(Li)), and revealed that the rotational 
velocity ($v_{\rm e}\sin i$) is the most influential key parameter
(Takeda et al. 2007; hereinafter referred to as Paper I).
In a successive study, we investigated the activities of these solar 
analogs by using the residual flux at the core of the Ca~{\sc ii} 8542 
line ($r_{0}$(8542)), and confirmed a clear correlation between
$A$(Li), $v_{\rm e}\sin i$, and $r_{0}$(8542), as expected
(Takeda et al. 2010; hereinafter referred to as Paper II).
However, it turned out hard to discriminate the differences 
in $r_{0}$(8542) when the activity is as low as that of the Sun, 
since it tends to get settled at $\sim 0.2$ and does not serve 
as a sensitive indicator any more. Actually, test calculations
of non-LTE line formation suggested (cf. Appendix B in Paper II) 
that the core flux of the Ca~{\sc ii} 8498/8542/8662 triplet lines 
are rather inert to the chromospheric temperature rise in the upper 
atmosphere when the activity is low, but that for the Ca~{\sc ii} 
3934/3968 doublet (H+K lines) is still sensitive and thus more 
advantageous for detecting the low-level activity.
Then, we recently studied the Be abundances of these sample stars 
by using the Be~{\sc ii} 3131 line based on the near-UV spectra
obtained with Subaru/HDS (Takeda et al. 2011; hereinafter referred to 
as Paper III). Since these HDS spectra fortunately cover the Ca~{\sc ii} H+K 
lines in the violet region, we decided to reinvestigate the activities
of these 118 Sun-like stars by measuring the core-emission strength
of the Ca~{\sc ii} 3934 (K) line, in order to clarify the activity
status of our Sun in comparison with similar solar analogs, where 
special attention was given to correctly removing the background 
line profile computed from the solar photospheric model, while 
taking advantage of the fact that atmospheric parameters of all these 
targets are well established. The purpose of this paper is to report
the outcome of this investigation.

\section{Basic Observational Data}

\subsection{Target Sample}

We use the same targets (118 solar analogs) as used in Papers I--III,
which were selected by the criteria of having $B-V$ and $M_{V}$ values
sufficiently similar to those of the Sun ($|\Delta (B-V)| \ltsim$~0.2--0.3
and $|\Delta M_{V}| \ltsim 0.3$). See section 2 in Paper I for a 
detailed description about the sample selection. We also determined
the atmospheric parameters ($T_{\rm eff}$, $\log g$, $v_{\rm t}$,
and [Fe/H]) from the equivalent widths of Fe lines, and the stellar 
parameters ($M$ and $age$) by comparing the positions on the HR diagram 
with the theoretical evolutionary tracks (cf. section 3 in Paper I).
These parameters for most of the targets were actually confirmed to 
proximally distribute around the solar values (i.e.,
$|\Delta T_{\rm eff}| \ltsim 100$~K, $|\Delta \log g| \ltsim 0.1$~dex,
$|\Delta v_{\rm t}| \ltsim 0.2$~km~s$^{-1}$, 
$|\Delta{\rm [Fe/H]}|\ltsim 0.2$~dex,
$|\Delta M|\ltsim 0.1\;M_{\odot}$, 
and $|\Delta \log age| \ltsim 0.5$~dex; cf. figures 4 and 5 in Paper I).

However, the following characteristics regarding the relations between 
these parameters are to be noted, which we had better bear in mind 
in discussing the behavior of stellar activities.\\
--- Since the effect of a decreased metallicity on $B-V$ is compensated 
by a lowering of $T_{\rm eff}$, several outlier stars with appreciably 
lower $T_{\rm eff}$ as well as [Fe/H] 
($\Delta T_{\rm eff} \ltsim -200$~K and [Fe/H]~$\ltsim -0.4$~dex)
are included in our sample (cf. figure 4c in Paper I), 
such as HIP~26381, HIP~39506, HIP~40118, and HIP~113989; and they belong to 
the oldest group ($age \sim 10^{10}$~yr).\\
--- These parameters are not independent from each other and some
correlations appear to exist between specific combinations; such as 
$age$ vs. $T_{\rm eff}$ (lower $T_{\rm eff}$ stars tend to be older), 
$age$ vs. [Fe/H] (lower [Fe/H] stars tend be older), 
$M$ vs. $age$ (lower-mass stars tend to be older), and 
$v_{\rm t}$ vs. $T_{\rm eff}$ ($v_{\rm t}$ tends to decrease with 
a lowered $T_{\rm eff}$), as recognized from figure 5 or figure 10 
in Paper I, though the existence of outlier stars mentioned above 
partly plays a role in these tendencies.

It should also be remarked that the data for HIP~41484 given in Paper I
were incorrect, as reported in appendix A of Paper II, where the 
correct results derived from a reanalysis of this star are presented. 

\subsection{Observational Material}

The spectroscopic observations of these 118 solar analogs and Vesta 
(substitute for the Sun) were carried out on 2009 August 6, 
2009 November 27, 2010 February 4, and 2010 May 24 (Hawaii Standard 
Time), with the High Dispersion Spectrograph (HDS; Noguchi et al. 2002) 
placed at the Nasmyth platform of the 8.2-m Subaru Telescope
atop Mauna Kea, by which we obtained high-dispersion spectra covering 
$\sim$~3000--4600~$\rm\AA$ with a resolving power of $R \simeq 60000$. 
See section 2 of Paper III and electronic table E1 therein for
details of the observations and the data reduction, as well as 
the basic data of the spectra (e.g., observing date, exposure time, 
S/N ratio at $\lambda \sim 3131 \rm\AA$).

The counts of raw echelle spectra (including the effect of blaze 
function) around $\lambda \sim 3950 \rm\AA$ (the broad maximum between 
the two large depressions of Ca~{\sc ii} 3934 (K) and 3968 (H) lines) 
are typically about $\sim 15$ times as large as those at the UV region
of $\lambda \sim 3131 \rm\AA$, while the counts at core of the Ca~{\sc ii} 
K line at 3934~$\rm\AA$ is about $\sim 10\%$ of those at 
$\lambda \sim 3950 \rm\AA$. Therefore, the S/N ratio at the deep 
absorption core of the K line is not much different
from the value at $\lambda \sim 3131 \rm\AA$ given in electronic 
table E1 of Paper III; that is, on the order of S/N~$\sim 100$.

\section{Core Emission Measurement}

\subsection{K line of ionized calcium}

Most activity studies of solar-type stars so far based on 
the core emission strengths of Ca~{\sc ii} resonance lines 
appear to utilize both K (3934~$\rm\AA$) and H (3968~$\rm\AA$) lines, 
presumably due to the intention of reducing the systematic 
errors by averaging both two, since measurements tend to be done 
rather roughly by directly integrating raw spectra at the specified
wavelength regions.

In this investigation, however, we focus only on the former K line 
at 3933.66~$\rm\AA$, since (1) it is by two times stronger than 
the H line and thus comparatively more suitable as a probe of
the condition at the optically-thin chromospheric layer, and
(2) the latter Ca~{\sc ii} H line at 3968.47~$\rm\AA$ is blended 
with the Balmer line (H$\epsilon$ at 3970.07~$\rm\AA$) which
would make the situation more complicated in simulating
the photospheric profile to be subtracted.

Our spectra in the selected wavelength region (3930--3937~$\rm\AA$)
including the relevant Ca~{\sc ii}~K line are shown in
figures 1 (Vesta/Sun), 2 (all 118 stars including Vesta/Sun),
and 3 (all stars overplotted), where the continuum levels of
the observed spectra are so adjusted as to match the theoretical ones
as explained below. We can see from these figures that the strengths
of core emission considerably vary from star to star, while
that for the Sun is apparently weak.

\setcounter{figure}{0}
\begin{figure}
  \begin{center}
    \FigureFile(80mm,60mm){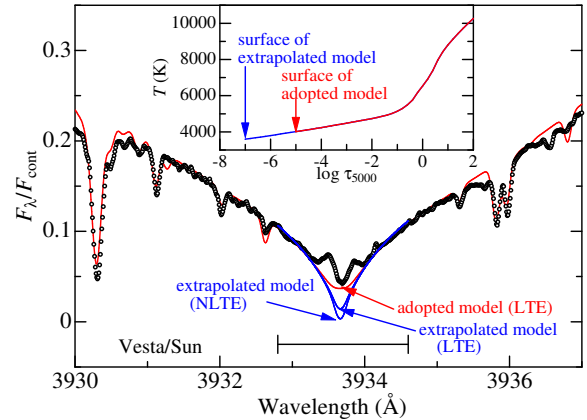}
  \end{center}
\caption{
Chromospheric emission feature at the core of the 
Ca~{\sc ii} 3933.663 line for the case of the Sun (Vesta),
in comparison with the theoretical photospheric spectra
computed from the classical model atmosphere. 
Three solid lines indicate the theoretical residual flux spectra 
(normalized by the continuum flux),
$r_{\lambda}^{\rm th} (\equiv F_{\lambda}^{\rm th}/F_{\rm cont}^{\rm th})$,
which are based on essentially the same Kurucz's (1993) 
ATLAS9 solar atmospheric model but with different surface locations 
(i.e., the optical depth at the first mesh point), as depicted in the 
$T(\tau_{5000})$ structures shown in the inset.
Red line is the LTE line profile derived from the adopted model 
atmosphere with its surface at $\log\tau_{5000} = -5$,
while blue lines are the LTE and NLTE line profiles derived from 
the specially-extrapolated model atmosphere with its surface at 
$\log\tau_{5000} = -7$.
Open symbols $\cdots$ observed spectrum of Vesta/Sun, 
$r_{\lambda}^{\rm obs} (\equiv D_{\lambda}^{\rm obs}/D_{\rm cont}^{\rm obs})$,
where $D_{\lambda}^{\rm obs}$ is the actually recorded spectrum
(ADU counts) while $D_{\rm cont}^{\rm obs}$ (regarded as an adjustable 
free parameter) was adequately chosen by requiring a satisfactory match 
between $r_{\lambda}^{\rm th}$ and $r_{\lambda}^{\rm obs}$
within $\ltsim$~2--3~$\rm\AA$ from the line center (excepting the 
core-emission region) as explained in subsection 3.2.
The specified line-core region (3932.8--3934.6~$\rm\AA$) is indicated
by a horizontal bar, over which the integration was made for evaluating 
the total chromospheric emission.
}
\end{figure}

\setcounter{figure}{2}
\begin{figure}
  \begin{center}
    \FigureFile(80mm,60mm){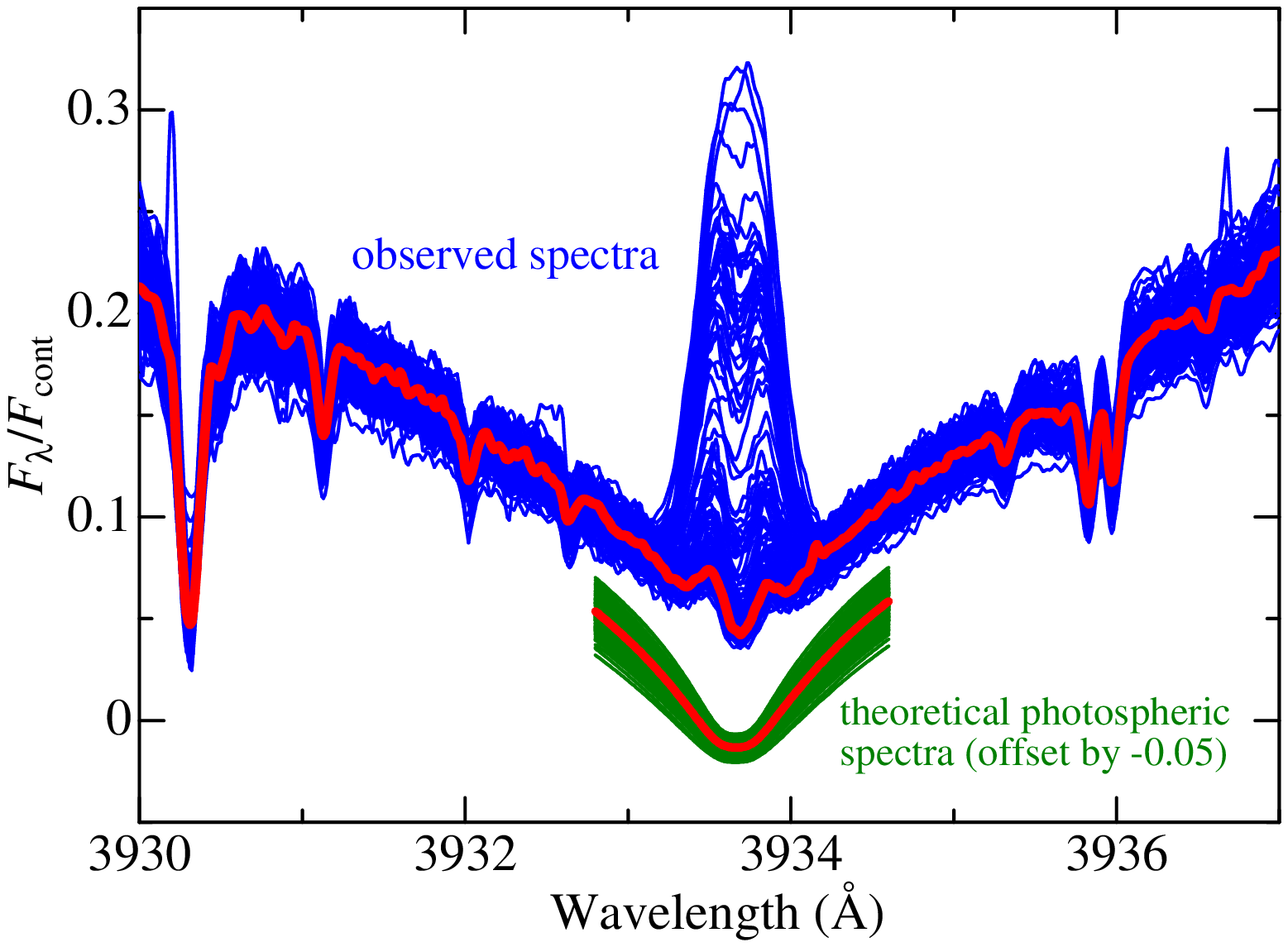}
  \end{center}
\caption{
In blue lines are overplotted the observed spectra 
($r_{\lambda}^{\rm obs}$) of Ca~{\sc ii} 3933.663 in the 
3930--3937~$\rm\AA$ region for all 118 stars (the same as those 
shown in figure 2), along with the Vesta/Sun spectrum highlighted 
in the red thick line. The theoretical photospheric spectra
($r_{\lambda}^{\rm th}$) for each of the stars are also overplotted 
in green lines (only the solar spectrum is in red) with a downward 
offset by $-0.05$. The wavelength scale of all spectra is adjusted 
to the laboratory frame.
}
\end{figure}

\subsection{Photospheric profile matching}

Given that the very strong Ca~{\sc ii} K and H lines with considerably 
extended damping wings are predominant at $\sim$~3900--4000~$\rm\AA$,
it is hopeless to empirically establish the continuum position
from the observed spectrum $D_{\lambda}^{\rm obs}$ (where an echelle 
order covers only $\sim 50$~$\rm\AA$).
We thus ``adjusted'' the continuum position ($D_{\rm cont}^{\rm obs}$) 
of the spectrum (judged by eye-inspection) in such a way that 
$r_{\lambda}^{\rm obs} (\equiv D_{\lambda}^{\rm obs}/D_{\rm cont}^{\rm obs})$
satisfactorily matches the theoretically calculated residual flux
$r_{\lambda}^{\rm th} (\equiv F_{\lambda}^{\rm th}/F_{\rm cont}^{\rm th})$\footnote{ 
We here use the astrophysical flux ($F$) defined by $\pi F_{\lambda}^{\rm th} 
\equiv 2\pi \int_{0}^{1} \mu I_{\lambda}^{\rm th}(0, \mu) d\mu$,
with which the effective temperature $T_{\rm eff}$ is related as
$\pi F_{\rm bol} = \pi \int_{0}^{\infty} F_{\lambda} d\lambda = \sigma T_{\rm eff}^{4}$
($\sigma$: Stephan--Boltzmann constant).}
in the inner wing of the line (within $|\Delta\lambda| \ltsim$~2--3~$\rm\AA$ from 
the line center, excepting the core-emission region).
An example of such an accomplished match is displayed in figure 1
for the case of the Sun (Vesta).

Regarding the computation of theoretical spectra, we used 
Kurucz's (1993) WIDTH9 program, which was modified by Y. Takeda 
to enable spectrum synthesis by including many lines.
As to the atomic line data, we invoked Kurucz and Bell's (1995)
compilation and included all available lines in the relevant region.
In particular, the data for the Ca~{\sc ii} K line at 3933.663~$\rm\AA$
of our primary concern are as follows: $\chi_{\rm low}$ = 0.00~eV, 
$\log gf = +0.134$, $\log \Gamma_{\rm R} = 8.20$ [radiation
damping width (s$^{-1}$)], 
$\log \Gamma_{e}/N_{\rm e} = -5.52$
[Stark effect damping width (s$^{-1}$) per electron density 
(cm$^{-3}$) at $10^{4}$~K], and 
$\log \Gamma_{w}/N_{\rm H} = -7.80$
[van der Waals damping width (s$^{-1}$) per hydrogen density 
(cm$^{-3}$) at $10^{4}$~K].
Since the atmospheric parameters ($T_{\rm eff}$, $\log g$, [Fe/H], and 
$v_{\rm t}$) are already established (as summarized in table 1), 
the model atmosphere for each star 
was generated by 3-dimensionally interpolating Kurucz's (1993) ATLAS9 model grids 
(LTE, plane-parallel model) in terms of $T_{\rm eff}$, $\log g$, and [Fe/H]. 
Then, the synthetic spectrum was computed by using the relevant atmospheric model
along with the metallicity-scaled abundances (for all elements including Ca; 
i.e., [X/H] = [Fe/H] for any X) as well as the microturbulence ($v_{\rm t}$),
and further broadened according to the macrobroadening parameter 
determined in Paper III. 

\subsection{Evaluation of chromospheric emission}

Now that the theoretical photospheric background profile ($r_{\lambda}^{\rm th}$) 
used for subtraction has been successfully fitted with $r_{\lambda}^{\rm obs}$ 
by appropriately adjusting the continuum position, we can calculate the absolute 
emission flux ($F'_{\rm Kp}$) at the K-line core originating from the chromosphere 
(i.e., after subtraction of the photospheric component) as 
\begin{equation}
F'_{\rm Kp} \equiv 
F_{\rm cont}^{\rm th}\int _{\lambda_{1}}^{\lambda_{2}}
(r_{\lambda}^{\rm obs} - r_{\lambda}^{\rm th})d \lambda
\end{equation}
where $\lambda_{1}$ and $\lambda_{2}$ defining the integration range 
were chosen to be 3932.8 and 3934.6~$\rm\AA$, respectively (cf. figure 1).
In order to demonstrate how this subtraction process works well, 
the photospheric profile ($r_{\lambda}^{\rm th}$) 
at [$\lambda_{1}$, $\lambda_{2}$] is displayed by red dashed line 
(along with the observed spectrum shown by symbols) for each star in figure 2. 
Finally, we can obtain $R'_{\rm Kp}$ (the ratio of the chromospheric 
emission flux at the K line to the total bolometric flux) as
\begin{equation}
R'_{\rm Kp} \equiv F'_{\rm Kp}/ F_{\rm bol} 
= \pi F'_{\rm Kp} / (\sigma T_{\rm eff}^{4}).
\end{equation}
The resulting $\log R'_{\rm Kp}$ values for each of the 118 solar analogs 
(+Sun) are presented in table 1, where other activity-related quantities 
($r_{0}$(8542) and $v_{\rm e}\sin i$) determined in Paper II are also 
given, along with the Li/Be abundances and stellar parameters established 
in Papers I and III. 

\subsection{Zero-Point Uncertainties in $R'_{Kp}$}

We would like to remark here that the ``absolute'' values of 
$R'_{\rm Kp}$ are not very meaningful in the low-activity regime
because of the uncertainties in its zero-point.
That is, the theoretical profile we have computed for subtraction of
the background photospheric component is by no means uniquely defined.
Actually, apparently different results may be obtained depending on 
how it is calculated.

This situation is demonstrated in figure 1 for the solar case.
The solar atmospheric model (which was obtained by interpolating 
the grids of Kurucz's ATLAS9 model atmospheres) we adopted for 
calculating the LTE photospheric profile (red line) has its surface 
at $\log \tau_{5000}^{\rm surf} = -5$ ($T^{\rm surf} \sim 4000$~K).
However, we note that the same but simply extrapolated model up to 
$\log \tau_{5000}^{\rm surf} = -7$ ($T^{\rm surf} \sim 3600$~K)
yields an appreciably deeper core (upper blue line), reflecting 
the fact that the residual flux at the center is determined by
$\sim B_{\lambda}(T^{\rm surf})/B_{\lambda}(T^{\rm ph})$
($T^{\rm ph}$: photospheric temperature) and very sensitive to
$T^{\rm surf}$ in this violet region where Wien's approximation
nearly holds. Moreover, the core of the non-LTE profile (simulated 
by using the departure coefficients computed in Paper II
for the case of Model E; cf. Appendix B therein) gets even more
deeper approaching a completely dark core (lower blue line).
Besides, the core shape strongly depends on the turbulent
velocity field in the upper atmosphere; e.g., compare the non-LTE 
profile in figure 1 (computed with a depth-independent microturbulence
of 1~km~s$^{-1}$) with that for Model E depicted in figure B.1(b)
of Paper II (where a variable microturbulent velocity field increasing
with height was adopted).

Accordingly, $R'_{\rm Kp}$ values [equation (2)] may 
be uncertain by an arbitrary constant, because 
$\int_{\lambda_{1}}^{\lambda_{2}}r_{\lambda}^{\rm th}d\lambda$
in equation (1) can have different values depending on
how $r_{\lambda}^{\rm th}$ is computed, though ``relative'' differences 
between $R'_{\rm Kp}$ values of each star are surely meaningful 
as long as they are evaluated in the same system.
Therefore, we should keep in mind that comparison of absolute 
values of our $R'_{\rm Kp}$ with those of similar $R'$ parameters 
derived by other groups, which we will try in subsection 4.1,
is not much meaningful when the low-activity region
(e.g., $\log R'_{\rm Kp} \ltsim -5$) is concerned. (On the other hand,
such a zero-point problem should not be so serious for higher-activity 
cases, where $\log R'_{\rm Kp}$ tends to be dominated by the emission 
component and the role of $r_{\lambda}^{\rm th}$ subtraction is 
less significant.) In any event, we consider that our choice of
shallower $r_{\lambda}^{\rm th}$ is adequate, since it leads to
smaller values of $R'_{\rm Kp}$ (as a result of larger subtraction), 
which eventually realizes a larger contrast in the $\log R'_{\rm Kp}$ 
values of low-activity stars. 

\section{Discussion}

\subsection{Comparison of $R'_{\rm Kp}$ with previous studies}

The $\log R'_{\rm Kp}$ values determined in subsection 3.3 are compared
with the equivalent activity indices\footnote{
Since Strassmeier et al. (2000) treated H and K lines separately
and presented each data of $F'_{\rm H}$ and $F'_{\rm K}$, we could convert 
their $R'_{\rm HK} [\equiv (F'_{\rm H} + F'_{\rm K})/F_{\rm bol}]$ 
(as clearly defined by them)
into $R'_{\rm K} (\equiv F'_{\rm K}/F_{\rm bol})$ which is directly 
comparable with our $R'_{\rm Kp}$. Meanwhile, the $R'_{\rm HK}$ data 
derived by Wright et al.'s (2004) as well as Isaacson and Fischer (2010)
appear to be essentially the {\it average} 
(not the sum) of $R'_{\rm H}$ and $R'_{\rm K}$ as judged by their
extents. Therefore, we should be cautious about different definitions 
in the meaning of $R'_{\rm HK}$. This situation is manifestly displayed
in figure 7 of Paper II, where we can see that $R'_{\rm HK}$(Strassmeier)
is systematically larger than $R'_{\rm HK}$(Wright) by 0.3~dex.
Therefore, we compared our $\log R'_{\rm Kp}$ with 
$R'_{\rm HK}$(Wright, Isaacson) without applying any correction, 
since the latter is practically 
equivalent to $R'_{\rm K}$ in any case.} derived by Strassmeier 
et al. (2000) [$\log R'_{\rm K}$], Wright et al. (2004) 
[$\log R'_{\rm HK}$], and Isaacson and Fischer (2010)
[$\log R'_{\rm HK}$] in figures 4a, 4b, and 4c, respectively.
These figures reveal almost the same tendency of correlation 
between our $\log R'$ and those of three studies: The agreement
at $\log R' \gtsim -5$ is mostly good, though our $\log R'$
tends to be slightly larger by $\sim 0.1$~dex for
high-activity stars of $\log R' \gtsim -4.5$. Meanwhile,
a distinct difference is observed at the low-activity region
where their $\log R'$ values tend to settle down at $\sim -5$, 
while ours are dispersed over $-5.5 \ltsim \log R' \ltsim -5$.
Admittedly, there is not much meaning in comparing the absolute 
$R'$ values of different systems with each other, as remarked 
in subsection 3.4. The important point is, however, that we 
could detect the subtle difference
in the low-level activity by measuring the weak core-emission 
at Ca~{\sc ii} K with a careful subtraction of the background 
photospheric profile, while such a precision for distinguishing 
the delicate difference in the weak emission strength could not 
be accomplished by comparatively rough measurements in those 
previous studies.

\setcounter{figure}{3}
\begin{figure}
  \begin{center}
    \FigureFile(70mm,140mm){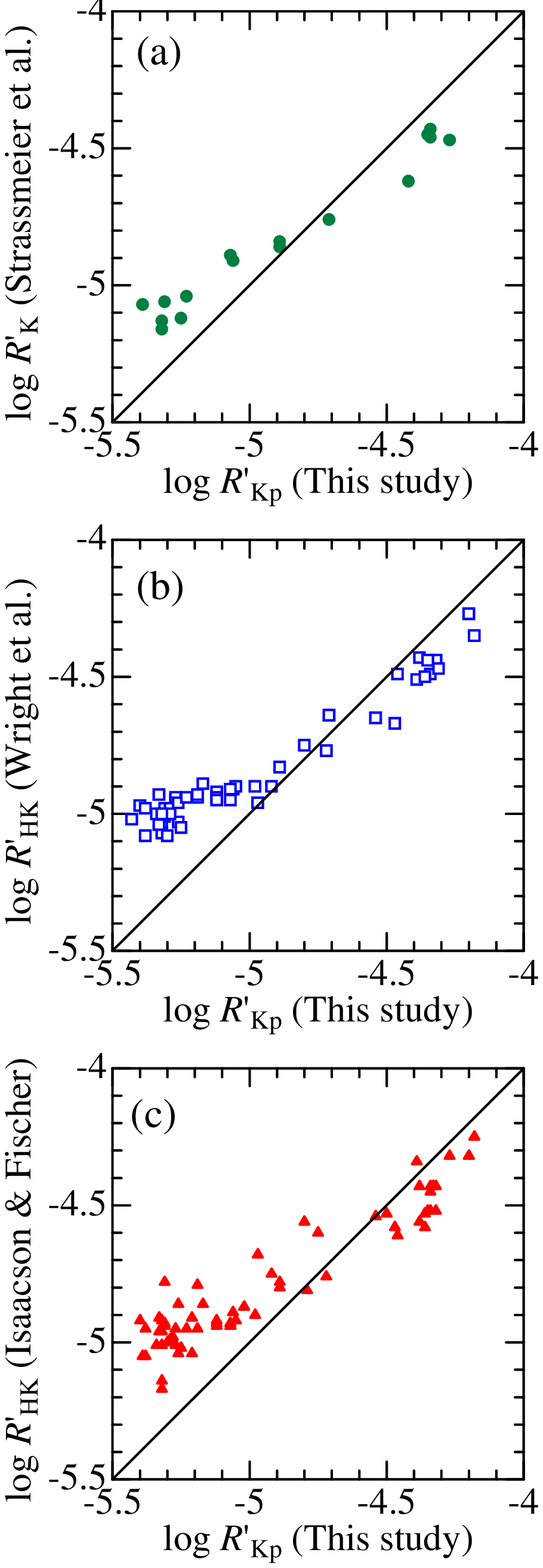}
  \end{center}
\caption{
Correlation of the $\log R'_{\rm Kp}$ indices determined 
in this study with the literature values taken from three 
representative papers:
(a) Strassmeier et al.'s (2000) $\log R'_{\rm K}$ values,
where 16 stars are in common. (b) Wright et al.'s (2004) 
$\log R'_{\rm HK}$ values, where 50 stars are in common.
(c) Isaacson and Fischer's (2010) $\log R'_{\rm HK}$ values, 
where 66 stars are in common.
(Since $\log R'_{\rm HK}$ of Wright et al. (2004) as well as 
Isaacson and Fischer (2010) should be equivalent to the mean 
of $\log R'_{\rm H}$ and $\log R'_{\rm K}$, it may be directly 
compared with our $\log R'_{\rm Kp}$.) 
}
\end{figure}

\subsection{Connection with stellar parameters}

Figures 5a, 5b, and 5c display how the three activity-related
parameters ($r_{0}$(8542), $v_{\rm e}\sin i$, and $\log age$;
cf. Paper II) are correlated with $\log R'_{\rm Kp}$.

\setcounter{figure}{4}
\begin{figure}
  \begin{center}
    \FigureFile(70mm,140mm){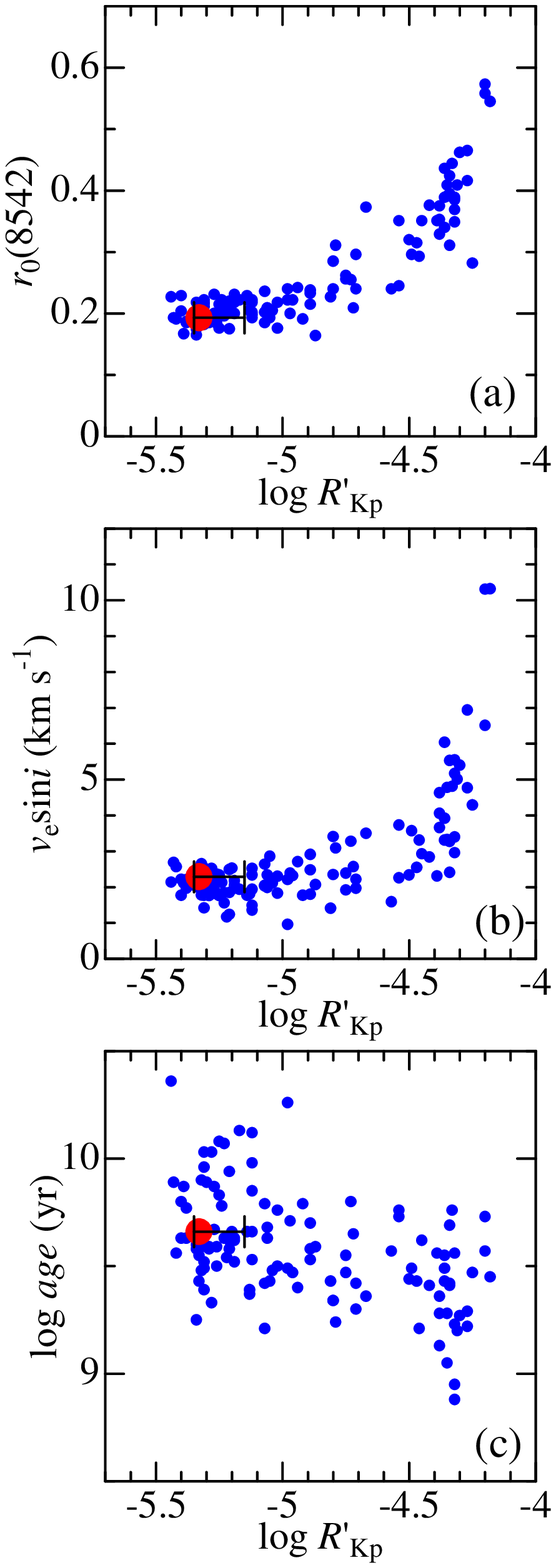}
  \end{center}
\caption{
Diagrams showing how the activity-related quantities
derived in Paper I and Paper II ($r_{0}(8542)$, $v_{\rm e}\sin i$,
and $\log age$; also given in table 1) are correlated with 
the activity index ($\log R'_{\rm Kp}$) derived in this study.
(a) $r_{0}(8542)$ vs. $\log R'_{\rm Kp}$,
(b) $v_{\rm e}\sin i$ vs. $\log R'_{\rm Kp}$, and 
(c) $\log age$ vs. $\log R'_{\rm Kp}$.
The $\log R'_{\rm Kp,\odot}$ value which we derived from the spectrum of
Vesta/Sun (corresponding to the near-minimum phase of activity) is indicated 
by the bigger (red) circle, while the expected minimum--maximum
range of $\log R'_{\rm Kp,\odot}$ (between $-5.35$ and $-5.15$;
cf. subsection 4.3) is shown by a horizontal bar.
}
\end{figure}

We can observe in figure 5a a marked sensitivity-difference
between $\log R'_{\rm Kp}$ and $r_{0}$(8542) in the low-activity 
region; i.e., the latter is inert to a variation of low-level 
activity and stabilizes at $\sim 0.2$, despite that the former 
still shows an appreciable variability over 
$-5.5 \ltsim \log R'_{\rm Kp} \ltsim -5$.
This is just what we have expected (cf. Appendix B in Paper II),
and demonstrates the superiority of the Ca~{\sc ii} HK core emission
(as long as correctly measured) to the line-center residual flux 
of Ca~{\sc ii} 8542 when it comes to investigating the activities 
of solar-type stars as low as the Sun.

Figure 5b shows a positive correlation between $R'_{\rm Kp}$ 
and $v_{\rm e}\sin i$, suggesting that stellar activity depends
on the rotation rate, as we already confirmed in Paper II
(cf. figure 5a therein). However, since $v_{\rm e}\sin i$ values
cluster around $\sim 2$~km~s$^{-1}$ at 
$-5.5 \ltsim \log R'_{\rm Kp} \ltsim -5$, we can not state much 
about whether this activity--rotation connection persists down to
such a low-activity region (also, uncertainties in the projection 
factor prevent from a meaningful discussion).

When we compare the $age$ vs. $R'_{\rm Kp}$ relation depicted in 
figure 5c with the similar $r_{0}$(8542) vs. $age$ plot 
(cf. figure 5c of Paper II), the anti-correlation is more clearly
(or less unambiguously, to say the least) recognized in the present 
case, thanks to the extended dynamic range of the activity indicator 
for low-activity stars, though the dispersion is still considerably
large. 

How the abundances of Li and Be depend on $R'_{\rm Kp}$ determined
in this study is illustrated in figure 6, where their dependences
upon $r_{0}$(8542) and $v_{\rm e}\sin i$ already discussed in Papers
II and III are also shown for comparison. We can see from
figures 6a that the near-linear relation between $A$(Li) and 
$\log R'_{\rm Kp}$ ($A$(Li)~$\simeq 7 + \log R'_{\rm Kp}$) holds 
widely from highly active ($\log R'_{\rm Kp} \sim -4$) to less 
active ($\log R'_{\rm Kp} \sim -5.5$) stars, 
in contrast to the case of $A$(Li) vs. $r_{0}$(8542) where
$A$(Li) shows a considerable dispersion at $r_{0}$(8542)~$\sim 0.2$
(as if compressed) because of the less sensitivity of $r_{0}$(8542).
This substantiates the observational conclusion in Paper II (or 
corroborates its validity even for less-active cases) that $A$(Li)
closely depends upon the stellar activity, which further lends 
support for our previous argument that the key parameter for controlling 
the surface lithium in solar-analog stars is the stellar rotation.

Regarding $A$(Be), we suspected in Paper III that the peculiar 4 stars
showing drastically depleted Be (by $\gtsim 2$~dex) in comparison with 
the solar abundance (while others have more or less near-solar Be) 
might be very slowly-rotating and and thus less active star than 
the Sun. However, since these 4 stars have appreciably different 
$\log R'_{\rm Kp}$ from each other
($-5.4 \ltsim \log R'_{\rm Kp} \ltsim -4.9$; cf. figure 6a$'$) 
and are not necessarily less active than the Sun, this speculation
does not seem likely. Some other explanation would have to be sought for.

\setcounter{figure}{5}
\begin{figure}
  \begin{center}
    \FigureFile(80mm,160mm){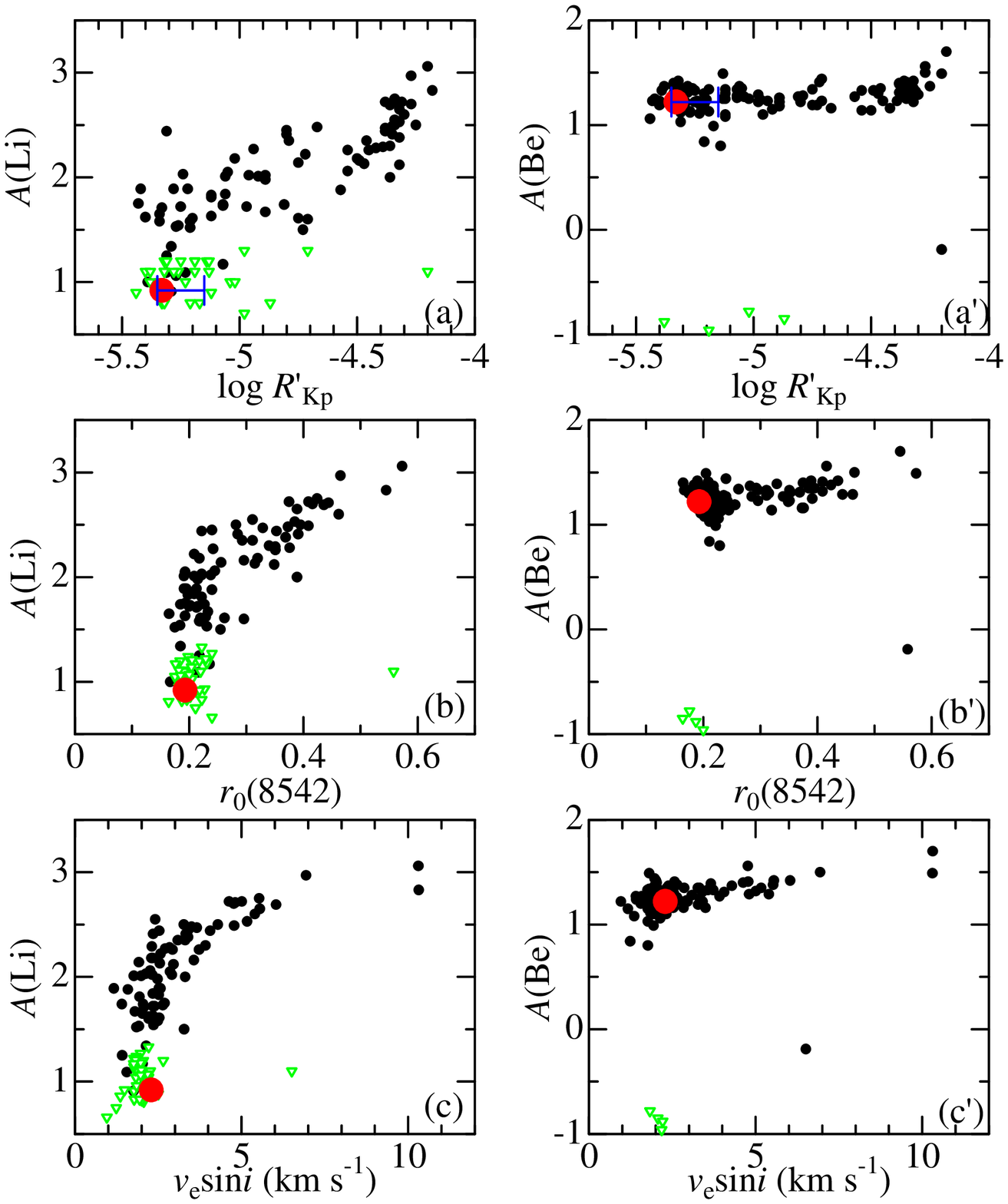}
  \end{center}
\caption{
Abundances of Li (left panels) and Be (right panels) plotted 
against $\log R'_{\rm Kp}$ (top), $r_{0}$(8542)
(middle), and $v_{\rm e}\sin i$ (bottom).
The solar values are indicated by the bigger (red) circle
in each panel. The expected minimum--maximum range of
$\log R'_{\rm Kp,\odot}$ is shown by a horizontal bar.
The results for the determinable cases are shown by filled circles, 
while the open inverse triangles denote the upper limits for 
the unmeasurable cases.
}
\end{figure}

\subsection{Activities of solar analogs and the Sun}

We now discuss the activity trends of 118 solar analogs, 
with special attention paid to the status of our Sun among 
its close associates, as the main subject of this study.
While our discussion is based on the $\log R'_{\rm Kp}$ indices
determined by ourselves, we should keep in mind  
that only one snapshot data is available for each star.
This means that uncertainties due to possible time-variations of stellar
activities (whichever cyclic or irregular) are inevitably involved.
For example, the Mt. Wilson $S$ value for the Sun varies over 
the range of $0.16 \ltsim S_{\odot} \ltsim 0.20$ (Baliunas et al. 1995),
which may be translated into a variability in $\log R'_{\odot}$ by 
$\sim 0.2$~dex ($-5.0 \ltsim \log R'_{\odot} \ltsim -4.8$)
according to the transformation formula given by Noyes et al. (1984). 
Since the observation time (2010 February 5) of our solar spectrum 
(Vesta) had better be regarded as corresponding to the near-minimum 
phase because of the appreciably retarded beginning of cycle 24 
after the minimum in 2008, our $\log R'_{\rm Kp,\odot} (= -5.33)$
may as well be raised by $\ltsim 0.2$~dex at the solar maximum phase. 
It is thus reasonable to assume that our $\log R'_{\rm Kp,\odot}$ 
ranges from $-5.35$ (solar minimum) to $-5.15$ (solar maximum),
as indicated by a short horizontal bar in figures 5, 6, and 7.

The distribution histogram of $\log R'_{\rm Kp}$ for all the 118 stars 
(and the Sun) and a similar histogram of $r_{0}$(8542) (for comparison; 
cf. Paper II) are shown in figures 7a and 7b, respectively.
We immediately notice in figure 7a the bimodal distribution of 
$\log R'_{\rm Kp}$ having two peaks at $\sim -5.3$ and $\sim -4.3$,
constituting a well-known Vaughan--Preston gap (Vaughan \& Preston 1980),
while this bimodal trend is not clear in the distribution of $r_{0}$(8542) 
(figure 7b) because of the densely peaked population around  
$r_{0}$(8542)~$\sim 0.2$, reflecting its insensitivity when the activity 
is low (cf. figure 5a).

As we can see from figure 7a, the Sun with $\log R'_{\rm Kp,\odot}$
of $-5.33$ manifestly belongs to the low-activity group (ranging 
from $\sim -5.4$ to $\sim -5.0$). As a matter of fact, only 11 
($\sim 10\%$)\footnote{
This ratio naturally depends on the reference solar activity,
which may be subject to uncertainties due to cyclic variations
as mentioned at the beginning of subsection 4.3. For example, if we assume 
a somewhat higher value of $-5.2$ for $\log R'_{\rm Kp,\odot}$ 
corresponding to the active phase of the Sun, this fraction 
becomes $\sim 30\%$.} 
out of 118 solar analogs have $\log R'_{\rm Kp}$ values smaller than $-5.33$. 
This distinctly low-activity nature of the Sun is also recognized 
by eye-inspection of figure 2, revealing that the Ca~{\sc ii} K line 
emission strength in our solar spectrum is near to the minimum level 
among other stars.

\setcounter{figure}{6}
\begin{figure}
  \begin{center}
    \FigureFile(70mm,120mm){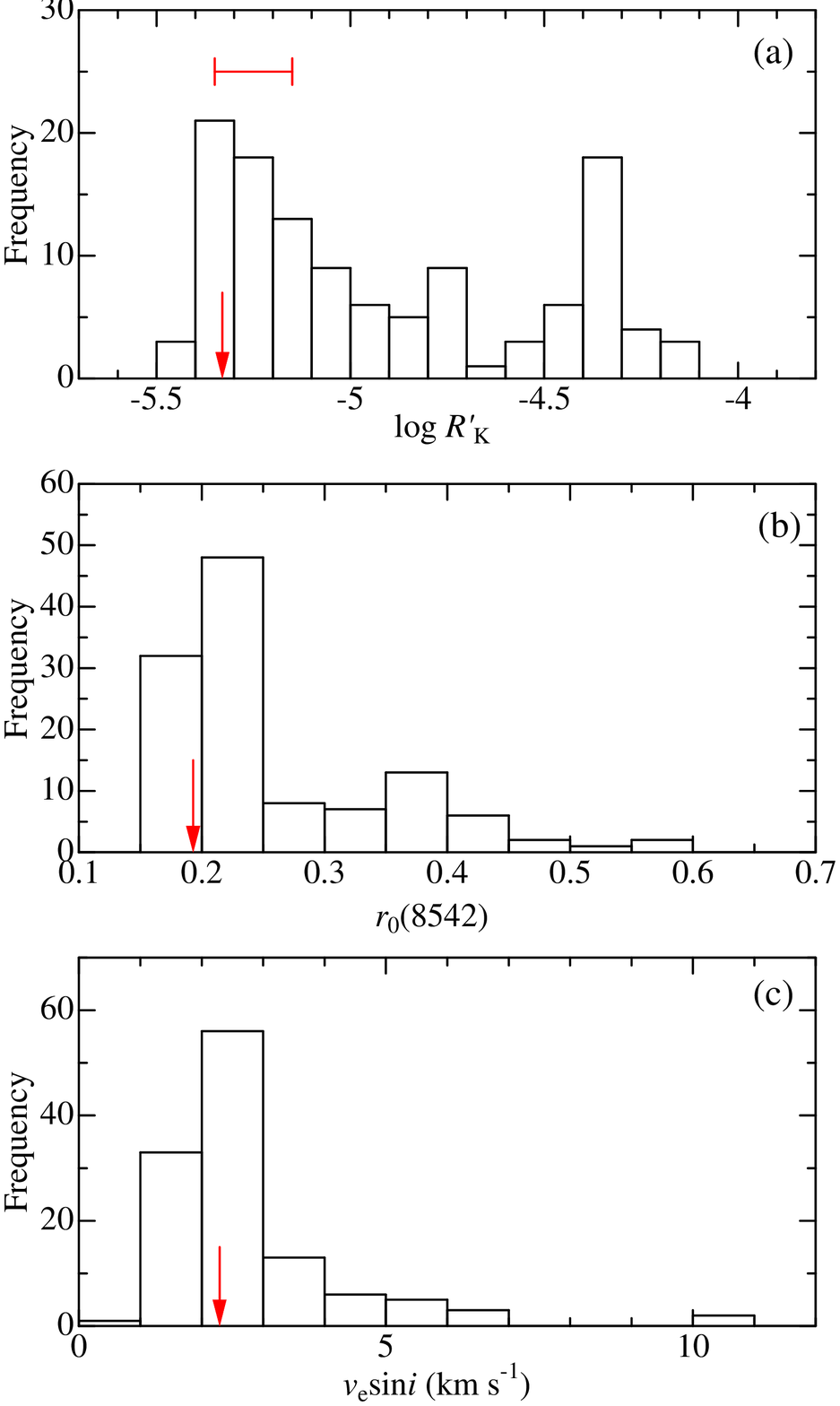}
  \end{center}
\caption{
Histograms showing the distributions of (a) $\log R'_{\rm Kp}$ 
and (b) $r_{0}(8542)$ for our sample of 118 solar analogs. 
The position for the solar value is indicated by an downward arrow 
at each panel. The expected minimum--maximum range of
$\log R'_{\rm Kp,\odot}$ is shown by a horizontal bar.
}
\end{figure}

Thus, we can rule out the possibility for the existence of a significant 
fraction of Maunder-minimum stars (i.e., solar-type stars with appreciably 
lower activity than the Sun, even showing an another peak well below 
the current solar-minimum level), such as those once suggested by 
Baliunas and Jastrow (1990). 
This result corroborates the arguments raised by recent studies 
(e.g., Hall \& Lockwood 2004; Wright 2004), which cast doubts about 
the reality of such a high frequency of Maunder-minimum stars.
Thus, our Sun belongs to the group of manifestly low activity 
level among solar analogs, the fraction of stars below which 
is essentially insignificant. 

\subsection{Stars of Subsolar Activity}

Although we have concluded that the Sun belongs to nearly the lowest
activity group, some stars do exist showing activities
still lower than that of the minimum-Sun, which are worth being 
examined more in detail. 
Since we defined that solar $\log R'_{\rm Kp, \odot}$ varies
from $-5.35$ (minimum) to $-5.15$ (maximum) (cf. section 4.1), 
such stars may be sorted out by the criterion of $\log R'_{\rm Kp} < -5.35$, 
which resulted in the following 8 objects ($\log R'_{\rm Kp}$,
$A$(Li), $\log age$, [Fe/H], remark):
HIP~7918 ($-5.42$, 1.89,  9.56, +0.01),
HIP~31965 ($-5.39$, 1.00,  9.87, +0.05),
HIP~39506 ($-5.44$, $<0.9$, 10.36, $-0.62$) 
HIP~53721 ($-5.43$, 1.75,  9.89, $-0.02$, PHS),
HIP~59610 ($-5.40$, 1.62,  9.63, $-0.06$, PHS),
HIP~64150 ($-5.38$, $<1.0$,  9.63, +0.05, Be depleted),
HIP~64747 ($-5.40$, $<1.1$,  9.80, $-0.18$), and
HIP~96901 ($-5.38$, $<1.1$,  9.77, +0.08, PHS). 

While only one (HIP~39506) of these is an outlier of lower 
[Fe/H] as well as lower $T_{\rm eff}$ (belonging to rather old 
population) as mentioned in subsection 2.1, the remaining 7 stars 
are Sun-like stars with sufficiently similar parameters,
which excludes the possibility of such a low-level activity being 
due to stellar-evolution (i.e., evolved subgiants; cf. section 1). 

We note here that (1) three (out of only five in our sample of 
118 stars) planet-host stars  are included,  (2) all these stars
have low-scale Li abundances ($A$(Li)~$\ltsim 2$), and (3) their ages 
are similar to or older than that of the Sun. Combining these facts with
the consequences in Papers I and II, we consider that these stars have
actually low activities even compared with the solar-minimum level,
and this is presumably attributed to their intrinsically slow rotation 
(which is closely related with the Li abundance as well as with 
the existence of giant planets). 

It is, therefore, interesting to investigate the variabilities
of these low-activity stars by long-term monitoring observations,
in order to see how their activities behave with time (cyclic? flat? 
irregular?). Admittedly, activity observations for these stars 
have been reported in several published studies so far, for example:
HIP~7918, 53721, 64150, and 96901 by Duncan et al. (1991);
HIP~53721, 59610, 64150, and 96901 by Wright et al. (2004);
HIP~7918, 53721, and 96901 by Hall et al. (2007);
HIP~31965, 59610, 64150, and 96901 by Isaacson and Fischer (2010).
However, as they are still quantitatively insufficient for establishing
the long-term behavior of their activities\footnote{As a comparatively 
well studied case where a sufficient amount of data are available, we may 
presumably state that HIP~96901 = HD~186427 showed a Sun-like cyclic 
variation in the 1995--2007 period; cf. figure 7 of Hall et al. (2007).},
much more observations are evidently needed.

\section{Conclusion}

There have been several arguments regarding the status of solar 
activity among similar Sun-like stars. which began with the
implication of Baliunas and Jastrow (1990) based on their Mt. Wilson
HK survey project that a considerable portion ($\sim 1/3$) of 
solar-type stars have activities significantly lower than
the present-day Sun, which they called ``Maunder-minimum stars.''
However, their conclusion could not be confirmed by Hall and Lockwood's
(2004) follow-up study, and Wright (2004) criticized the reality
of such considerably low-active solar-type stars by pointing out 
that most of them are not so much dwarfs as evolved subgiants.

Given this controversial situation, we decided to contend with 
this problem by ourselves based on carefully selected sample of 
118 solar-analogs sufficiently similar to each other (which we 
already investigated their stellar parameters as well as Li/Be 
abundances in a series of our previous papers), with a special 
attention being paid to reliably evaluating their activities 
down to a considerably low level. 

Practically, we measured the emission strength at the core of 
Ca~{\sc ii} 3933.663 line (K line) on the high-dispersion 
spectrogram obtained by Subaru/HDS, where we gave effort
to correctly evaluating the pure emission component by removing 
the wing-fitted photospheric profile calculated from 
the classical solar model atmosphere, which enabled us to detect 
low-level activities down to $\log R'_{\rm Kp} \sim -5.5$.

A comparison of our $\log R'_{\rm Kp}$ results with the corresponding 
$\log R'$ values of Strassmeier et al. (2000), Wright et al. (2004),
and Isaacson and Fischer (2010) 
revealed that low-active stars (for which they derived 
$\log R' \sim -5.1$ at the minimum limit) actually have a dispersion 
of $\sim 0.4$~dex ($-5.5 \ltsim \log R'_{\rm Kp} \ltsim -5.0$) 
in our measurement,
suggesting that our $\log R'_{\rm Kp}$ has a higher sensitivity
and thus advantageous. A similar situation holds regarding the 
comparison with $r_{0}$(8542) we used in Paper II; i,e., this index 
stabilizes at $\sim 0.2$ and becomes insensitive for low-active stars
in contrast to $\log R'_{\rm Kp}$. 

As another merit of using $\log R'_{\rm Kp}$, we can state
that the visibility of the $A$(Li)--activity relation as well as 
the age--activity relation becomes comparatively clearer, because
this activity index turns out to have well diversified values
for low-activity stars thanks to its high sensitivity, which 
can not be accomplished by using, e.g., $r_{0}$(8542).

From the distribution histogram of $\log R'_{\rm Kp}$, we could
recognize a clear Vaughan--Preston gap between two peaks
at $\sim -5.3$ and $\sim -4.3$.
Our result of $\log R'_{\rm Kp,\odot} = -5.33$ manifestly suggests 
that the Sun belongs to the group of the former peak and has a 
distinctly low-active nature among solar analogs. 
Actually, a fraction of stars with 
$\log R'_{\rm Kp} \le \log R'_{\rm Kp,\odot}$ is only $\sim 10\%$.
This consequence exclude the possibility for the existence
of a considerable fraction (e.g., $\sim 1/3$) of 
``Maunder-minimum stars'' such that having activities 
significantly lower than the current solar-minimum level as once 
suggested by Baliunas and Jastrow (1990).

Yet, some stars (only a minor fraction of the sample) do exist 
showing activities still lower than that of the solar-minimum level. 
Having examined such 8 low-activity stars, we found that they tend to
include planet-host stars and have low Li abundances, from which 
we suspect that their activities are actually low as a result of 
intrinsically slow rotation. It would be an important task to 
clarify the behavior of their activity variations by long-term 
monitoring observations.  


\clearpage

\onecolumn

\setcounter{figure}{1}
\begin{figure}
  \begin{center}
    \FigureFile(160mm,240mm){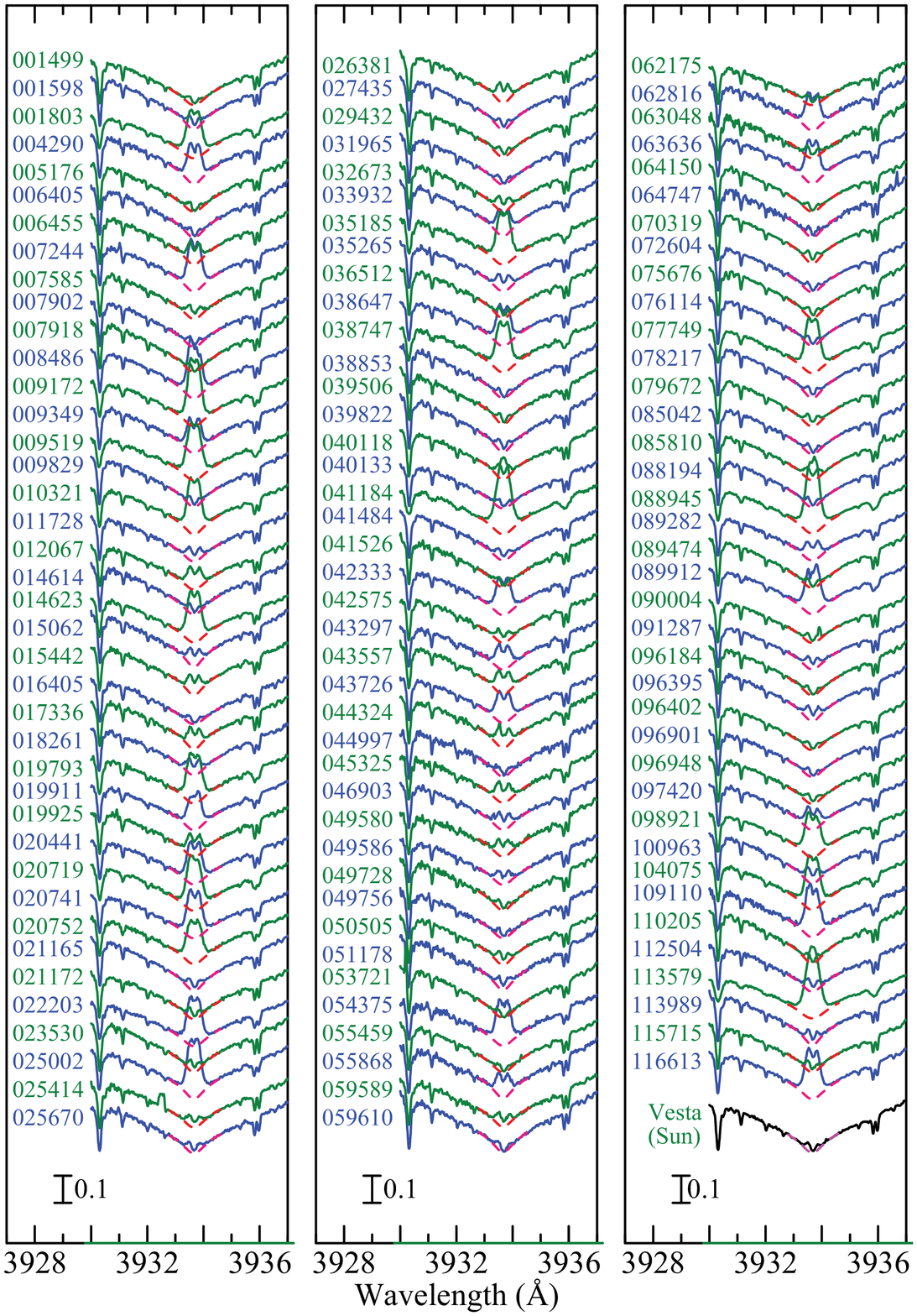}
  \end{center}
\caption{
Display of the 3930--3937~$\rm\AA$ region spectra ($r_{\lambda}^{\rm obs}$) 
of the Ca~{\sc ii} K line at 3933.663~$\rm\AA$ for all the 118 program stars 
along with the Vesta/Sun, where the theoretical line profile 
($r_{\lambda}^{\rm th}$)
calculated from the solar photospheric model is also overplotted by 
(red) dashed line in the core region (3932.8--3934.6~$\rm\AA$) around the
line center.
Each spectrum is vertically shifted by 0.1 (in continuum unit)
relative to the adjacent one.
The wavelength scale of all stellar spectra is adjusted to 
the laboratory frame by correcting the radial velocity
shifts. The HIP numbers are indicated in the figure. 
}
\end{figure}

\newpage

\setcounter{table}{0}
\small
\renewcommand{\arraystretch}{0.8}
\setlength{\tabcolsep}{3pt}
\begin{longtable}{cccrccrrccrl}
\caption{Activity index, rotation, Li abundance, age, and the atmospheric parameters.}
\hline\hline
HIP & $\log R'_{\rm Kp}$ & $r_{0}$(8542) & $v_{\rm e}\sin i$ & $A$(Li) & $A$(Be) 
& $\log age$ & $T_{\rm eff}$ & $\log g$ & $v_{\rm t}$ & [Fe/H] & Remark \\ 
\hline
\endhead
\hline
\endfoot
\hline
\multicolumn{12}{l}{\hbox to 0pt{\parbox{180mm}{\footnotesize
Note. 
Following the HIP number in column 1, 
$\log R'_{\rm Kp}$ (column 2) is the emission-strength index of the Ca~{\sc ii} K line 
determined in this study, $r_{0}(8542)$ (column 3) is the residual flux 
at the line center of the Ca~{\sc ii} 8542 line (cf. Paper II),
$v_{\rm e}\sin i$ (column 4) is the projected rotational velocity in km~s$^{-1}$ 
calibrated in Paper II, $A$(Li) (column 5) as well as 
$A$(Be) (column 6) are the lithium and beryllium abundances 
derived in Papers I and III (in the usual normalization of $A$(H) = 12.00), 
and $\log age$ (column 7) is the logarithm of the stellar age 
(in yr) derived from theoretical evolutionary tracks in Paper I.
In columns 8--11 are presented the atmospheric parameters for
each star, which are the ``standard parameters'' derived from Fe~{\sc i}
and Fe~{\sc ii} lines (cf. section 3.1.1 in Paper I): 
$T_{\rm eff}$ (effective temperature: in K) $\log g$ (surface gravity;
in cm~s$^{-2}$), $v_{\rm t}$ (microturbulence; in km~s$^{-1}$), and 
[Fe/H] (metallicity defined by $A$(Fe)$-7.50$, where $A$(Fe) is the 
logarithmic Fe abundance), respectively. In column 12 (remark), 
``PHS'' denotes planet-host stars.
}}}
\endlastfoot
\hline
001499 &  $-$5.32 &  0.190 &  2.01 &($<$1.1) &     1.42 &  9.59 &   5724  &  4.45  &  0.95 &  $+0.20$ & \\
001598 &  $-$5.12 &  0.222 &  1.94 &    1.81 &     1.10 &  9.98 &   5693  &  4.33  &  0.96 &  $-0.27$ & \\
001803 &  $-$4.32 &  0.389 &  5.55 &    2.65 &     1.42 &  8.88 &   5817  &  4.41  &  1.17 &  $+0.24$ & \\
004290 &  $-$4.50 &  0.320 &  2.34 &    2.18 &     1.14 &  9.44 &   5719  &  4.40  &  1.10 &  $-0.12$ & \\
005176 &  $-$5.28 &  0.195 &  2.52 &    1.89 &     1.37 &  9.33 &   5855  &  4.39  &  1.03 &  $+0.19$ & \\
006405 &  $-$5.33 &  0.213 &  2.07 &    1.71 &     1.24 &  9.55 &   5728  &  4.38  &  0.96 &  $-0.14$ & \\
006455 &  $-$4.81 &  0.227 &  1.41 &    1.74 &     1.27 &  9.43 &   5716  &  4.57  &  0.99 &  $-0.09$ & \\
007244 &  $-$4.39 &  0.351 &  2.31 &    2.29 &     1.23 &  9.56 &   5755  &  4.52  &  1.12 &  $-0.04$ & \\
007585 &  $-$5.06 &  0.209 &  2.34 &    1.84 &     1.35 &  9.68 &   5784  &  4.50  &  1.04 &  $+0.07$ & \\
007902 &  $-$5.31 &  0.182 &  2.13 &($<$0.9) &     1.28 &  9.39 &   5613  &  4.39  &  0.91 &  $-0.01$ & \\
007918 &  $-$5.42 &  0.191 &  2.57 &    1.89 &     1.25 &  9.56 &   5841  &  4.30  &  1.12 &  $+0.01$ & \\
008486 &  $-$4.34 &  0.311 &  2.41 &    2.55 &     1.33 &  9.69 &   5805  &  4.45  &  1.13 &  $-0.06$ & \\
009172 &  $-$4.25 &  0.282 &  4.29 &    2.50 &     1.37 &  9.47 &   5763  &  4.56  &  1.12 &  $+0.06$ & \\
009349 &  $-$4.54 &  0.245 &  2.26 &    2.06 &     1.14 &  9.73 &   5788  &  4.35  &  1.07 &  $+0.01$ & \\
009519 &  $-$4.27 &  0.465 &  6.94 &    2.97 &     1.50 &  9.29 &   5853  &  4.45  &  1.22 &  $+0.14$ & \\
009829 &  $-$5.31 &  0.211 &  1.77 &($<$0.9) &     1.03 & 10.03 &   5579  &  4.25  &  0.94 &  $-0.31$ & \\
010321 &  $-$4.34 &  0.395 &  3.27 &    2.50 &     1.35 &  9.41 &   5707  &  4.60  &  1.04 &  $-0.01$ & \\
011728 &  $-$5.04 &  0.205 &  2.11 &($<$1.0) &     1.27 &  9.48 &   5708  &  4.40  &  1.02 &  $+0.02$ & \\
012067 &  $-$4.71 &  0.240 &  1.97 &($<$1.3) &     1.44 &  9.42 &   5709  &  4.41  &  0.96 &  $+0.20$ & \\
014614 &  $-$5.21 &  0.220 &  2.49 &    1.58 &     1.20 &  9.58 &   5726  &  4.26  &  1.00 &  $-0.12$ & \\
014623 &  $-$4.36 &  0.389 &  3.31 &    2.00 &     1.35 &  9.49 &   5742  &  4.52  &  1.09 &  $+0.12$ & \\
015062 &  $-$4.96 &  0.222 &  2.32 &    2.02 &     1.10 &  9.47 &   5735  &  4.49  &  0.94 &  $-0.29$ & \\
015442 &  $-$4.89 &  0.233 &  1.80 &    1.67 &     1.18 &  9.70 &   5682  &  4.50  &  0.87 &  $-0.19$ & \\
016405 &  $-$5.32 &  0.184 &  2.65 &($<$1.2) &     1.38 &  9.48 &   5738  &  4.32  &  1.03 &  $+0.26$ & \\
017336 &  $-$4.87 &  0.164 &  2.07 &($<$0.8) &($<-0.9$) &  9.59 &   5671  &  4.55  &  0.94 &  $-0.13$ & Be depleted \\
018261 &  $-$4.94 &  0.242 &  2.71 &    2.27 &     1.23 &  9.40 &   5873  &  4.43  &  0.97 &  $+0.02$ & \\
019793 &  $-$4.32 &  0.385 &  5.17 &    2.53 &     1.35 &  9.23 &   5828  &  4.51  &  1.26 &  $+0.19$ & \\
019911 &  $-$4.54 &  0.351 &  3.73 &    2.26 &     1.33 &  9.76 &   5672  &  4.34  &  1.10 &  $-0.13$ & \\
019925 &  $-$4.75 &  0.262 &  2.39 &    1.61 &     1.34 &  9.47 &   5767  &  4.53  &  0.99 &  $+0.07$ & \\
020441 &  $-$4.42 &  0.376 &  2.84 &    2.28 &     1.16 &  9.41 &   5771  &  4.42  &  1.10 &  $+0.13$ & \\
020719 &  $-$4.30 &  0.462 &  5.40 &    2.60 &     1.29 &  9.27 &   5831  &  4.36  &  1.24 &  $+0.13$ & \\
020741 &  $-$4.35 &  0.391 &  3.33 &    2.41 &     1.25 &  9.05 &   5797  &  4.37  &  1.20 &  $+0.16$ & \\
020752 &  $-$4.38 &  0.375 &  4.63 &    2.72 &     1.40 &  9.28 &   5923  &  4.46  &  1.13 &  $+0.16$ & \\
021165 &  $-$5.20 &  0.220 &  2.53 &    1.61 &     1.24 &  9.66 &   5760  &  4.28  &  0.99 &  $-0.16$ & \\
021172 &  $-$5.23 &  0.196 &  1.85 &($<$1.0) &     1.11 &  9.63 &   5625  &  4.27  &  0.90 &  $-0.10$ & \\
022203 &  $-$4.36 &  0.340 &  3.92 &    2.30 &     1.27 &  9.55 &   5740  &  4.33  &  1.07 &  $+0.13$ & \\
023530 &  $-$5.21 &  0.211 &  1.24 &($<$0.8) &     0.84 &  9.94 &   5601  &  4.36  &  0.91 &  $-0.24$ & \\
025002 &  $-$4.32 &  0.369 &  3.40 &    2.38 &     1.31 &  9.56 &   5729  &  4.47  &  1.07 &  $-0.08$ & \\
025414 &  $-$5.13 &  0.220 &  1.85 &($<$1.1) &     1.25 &  9.37 &   5635  &  4.49  &  0.89 &  $+0.10$ & \\
025670 &  $-$5.19 &  0.218 &  2.05 &($<$1.2) &     1.34 &  9.52 &   5759  &  4.55  &  0.88 &  $+0.10$ & \\
026381 &  $-$4.98 &  0.240 &  0.96 &($<$0.7) &     1.22 & 10.26 &   5518  &  4.47  &  0.87 &  $-0.45$ & PHS, outlier in [Fe/H]/$T_{\rm eff}$ \\
027435 &  $-$5.27 &  0.231 &  1.92 &    1.53 &     1.13 &  9.87 &   5697  &  4.45  &  0.93 &  $-0.22$ & \\
029432 &  $-$5.27 &  0.200 &  2.14 &    1.06 &     1.12 &  9.67 &   5712  &  4.32  &  1.00 &  $-0.12$ & \\
031965 &  $-$5.39 &  0.167 &  2.11 &    1.00 &     1.33 &  9.87 &   5770  &  4.31  &  0.99 &  $+0.05$ & \\
032673 &  $-$5.02 &  0.176 &  1.83 &($<$1.0) &($<-0.8$) &  9.50 &   5724  &  4.57  &  0.95 &  $+0.06$ & Be depleted \\
033932 &  $-$4.67 &  0.373 &  3.50 &    2.48 &     1.16 &  9.36 &   5891  &  4.38  &  1.10 &  $-0.12$ & \\
035185 &  $-$4.33 &  0.444 &  4.81 &    2.71 &     1.29 &  9.76 &   5793  &  4.19  &  1.35 &  $ 0.00$ & \\
035265 &  $-$4.92 &  0.191 &  1.77 &    2.01 &     1.15 &  9.79 &   5804  &  4.37  &  1.04 &  $-0.02$ & \\
036512 &  $-$5.31 &  0.218 &  1.42 &    1.25 &     1.24 &  9.52 &   5718  &  4.49  &  0.89 &  $-0.09$ & \\
038647 &  $-$4.47 &  0.315 &  2.55 &    2.13 &     1.30 &  9.43 &   5714  &  4.43  &  0.95 &  $+0.01$ & \\
038747 &  $-$4.34 &  0.424 &  5.53 &    2.75 &     1.38 &  9.42 &   5804  &  4.42  &  1.05 &  $+0.07$ & \\
038853 &  $-$5.31 &  0.222 &  2.53 &    2.44 &     1.16 &  9.60 &   5899  &  4.27  &  1.03 &  $-0.05$ & \\
039506 &  $-$5.44 &  0.227 &  2.14 &($<$0.9) &     1.06 & 10.36 &   5600  &  4.24  &  0.83 &  $-0.62$ & outlier in [Fe/H]/$T_{\rm eff}$\\
039822 &  $-$5.19 &  0.231 &  1.97 &($<$1.2) &     1.13 &  9.63 &   5758  &  4.35  &  0.90 &  $-0.22$ & \\
040118 &  $-$5.12 &  0.202 &  1.36 &($<$0.9) &     1.08 & 10.12 &   5541  &  4.45  &  0.84 &  $-0.42$ & outlier in [Fe/H]/$T_{\rm eff}$\\
040133 &  $-$5.26 &  0.184 &  2.36 &    1.54 &     1.35 &  9.59 &   5698  &  4.33  &  0.97 &  $+0.12$ & \\
041184 &  $-$4.18 &  0.545 & 10.32 &    2.83 &     1.70 &  9.45 &   5705  &  4.43  &  1.51 &  $+0.11$ & \\
041484 &  $-$5.07 &  0.202 &  2.64 &    1.73 &     1.26 &  9.79 &   5864  &  4.33  &  0.92 &  $+0.05$ & \\
041526 &  $-$5.24 &  0.222 &  2.13 &    2.03 &     1.23 &  9.78 &   5801  &  4.27  &  0.98 &  $-0.02$ & \\
042333 &  $-$4.46 &  0.293 &  3.31 &    2.35 &     1.35 &  9.21 &   5816  &  4.44  &  1.08 &  $+0.14$ & \\
042575 &  $-$5.07 &  0.236 &  2.03 &    1.17 &     1.28 &  9.42 &   5675  &  4.40  &  0.96 &  $+0.06$ & \\
043297 &  $-$4.71 &  0.296 &  2.22 &    1.60 &     1.23 &  9.30 &   5691  &  4.46  &  1.05 &  $+0.08$ & \\
043557 &  $-$4.73 &  0.255 &  3.28 &    1.50 &     1.19 &  9.80 &   5805  &  4.42  &  1.05 &  $-0.06$ & \\
043726 &  $-$4.57 &  0.240 &  1.59 &    1.88 &     1.27 &  9.57 &   5769  &  4.48  &  1.01 &  $+0.11$ & \\
044324 &  $-$4.80 &  0.285 &  2.35 &    2.41 &     1.27 &  9.34 &   5888  &  4.46  &  1.09 &  $-0.01$ & \\
044997 &  $-$5.31 &  0.198 &  1.84 &($<$1.2) &     1.32 &  9.49 &   5696  &  4.54  &  0.75 &  $+0.04$ & \\
045325 &  $-$4.79 &  0.311 &  3.09 &    2.35 &     1.28 &  9.24 &   5935  &  4.47  &  0.97 &  $+0.18$ & \\
046903 &  $-$4.89 &  0.238 &  2.91 &    2.02 &     1.22 &  9.53 &   5746  &  4.40  &  1.11 &  $-0.03$ & \\
049580 &  $-$4.89 &  0.215 &  2.48 &    1.98 &     1.26 &  9.58 &   5782  &  4.41  &  0.87 &  $+0.02$ & \\
049586 &  $-$4.98 &  0.222 &  2.21 &($<$1.3) &     1.29 &  9.49 &   5786  &  4.42  &  1.06 &  $+0.20$ & \\
049728 &  $-$5.34 &  0.175 &  2.01 &($<$1.0) &     1.30 &  9.60 &   5744  &  4.40  &  0.98 &  $-0.07$ & \\
049756 &  $-$5.29 &  0.185 &  2.14 &    1.34 &     1.21 &  9.58 &   5720  &  4.29  &  0.99 &  $+0.02$ & \\
050505 &  $-$5.12 &  0.218 &  1.49 &($<$0.9) &     1.25 &  9.85 &   5590  &  4.44  &  0.84 &  $-0.17$ & \\
051178 &  $-$5.14 &  0.229 &  1.77 &($<$1.2) &     0.80 &  9.66 &   5801  &  4.47  &  0.87 &  $-0.17$ & \\
053721 &  $-$5.43 &  0.193 &  2.69 &    1.75 &     1.23 &  9.89 &   5819  &  4.19  &  1.15 &  $-0.02$ & PHS \\
054375 &  $-$4.38 &  0.353 &  4.06 &    2.44 &     1.31 &  9.36 &   5803  &  4.37  &  0.96 &  $+0.14$ & \\
055459 &  $-$5.34 &  0.218 &  2.39 &    1.58 &     1.37 &  9.58 &   5812  &  4.36  &  1.03 &  $+0.07$ & \\
055868 &  $-$4.75 &  0.256 &  1.92 &    2.14 &     1.19 &  9.55 &   5757  &  4.49  &  0.95 &  $-0.15$ & \\
059589 &  $-$5.13 &  0.205 &  1.81 &($<$1.2) &     1.49 &  9.39 &   5654  &  4.52  &  0.70 &  $-0.01$ & \\
059610 &  $-$5.40 &  0.229 &  2.22 &    1.62 &     1.23 &  9.63 &   5829  &  4.34  &  1.04 &  $-0.06$ & PHS \\
062175 &  $-$5.12 &  0.198 &  2.52 &    1.83 &     1.34 &  9.53 &   5683  &  4.19  &  0.90 &  $+0.13$ & \\
062816 &  $-$4.49 &  0.296 &  3.57 &    2.16 &     1.33 &  9.49 &   5804  &  4.44  &  0.97 &  $+0.06$ & \\
063048 &  $-$5.26 &  0.189 &  1.81 &($<$1.1) &     1.34 &  9.50 &   5655  &  4.32  &  0.91 &  $-0.02$ & \\
063636 &  $-$4.45 &  0.351 &  2.93 &    2.26 &     1.23 &  9.62 &   5799  &  4.52  &  1.10 &  $-0.01$ & \\
064150 &  $-$5.38 &  0.187 &  2.21 &($<$1.0) &($<-0.9$) &  9.63 &   5726  &  4.42  &  0.99 &  $+0.05$ & Be depleted \\
064747 &  $-$5.40 &  0.204 &  1.77 &($<$1.1) &     1.19 &  9.80 &   5710  &  4.42  &  0.93 &  $-0.18$ & \\
070319 &  $-$5.23 &  0.207 &  1.56 &    1.09 &     1.20 & 10.07 &   5678  &  4.42  &  0.96 &  $-0.33$ & \\
072604 &  $-$5.28 &  0.191 &  2.26 &($<$1.1) &     1.31 & 10.03 &   5655  &  4.24  &  0.84 &  $-0.14$ & \\
075676 &  $-$5.19 &  0.200 &  2.18 &($<$1.1) &($<-1.0$) &  9.62 &   5772  &  4.44  &  0.88 &  $-0.08$ & Be depleted \\
076114 &  $-$5.29 &  0.195 &  1.76 &    0.91 &     1.22 &  9.59 &   5709  &  4.42  &  1.02 &  $-0.02$ & \\
077749 &  $-$4.27 &  0.416 &  4.77 &    2.70 &     1.56 &  9.22 &   5836  &  4.61  &  1.14 &  $+0.22$ & \\
078217 &  $-$5.22 &  0.213 &  1.17 &    1.89 &     1.15 &  9.54 &   5749  &  4.43  &  1.10 &  $-0.22$ & \\
079672 &  $-$5.12 &  0.193 &  2.34 &    1.63 &     1.30 &  9.66 &   5768  &  4.40  &  0.96 &  $+0.04$ & \\
085042 &  $-$5.33 &  0.187 &  2.01 &($<$0.8) &     1.34 &  9.43 &   5676  &  4.48  &  0.99 &  $+0.03$ & \\
085810 &  $-$5.05 &  0.193 &  2.86 &    2.05 &     1.35 &  9.43 &   5856  &  4.46  &  1.08 &  $+0.15$ & \\
088194 &  $-$5.32 &  0.196 &  1.78 &($<$0.8) &     1.25 &  9.67 &   5693  &  4.33  &  0.98 &  $-0.08$ & \\
088945 &  $-$4.20 &  0.558 &  6.51 &($<$1.1) &  $-$0.19 &  9.73 &   5800  &  4.38  &  1.44 &  $-0.01$ & outlier in $A$(Li)\\
089282 &  $-$4.80 &  0.240 &  3.41 &    2.45 &     1.22 &  9.34 &   5833  &  4.22  &  1.00 &  $ 0.00$ & \\
089474 &  $-$5.30 &  0.189 &  2.50 &($<$0.9) &     1.25 &  9.89 &   5755  &  4.20  &  1.04 &  $+0.01$ & \\
089912 &  $-$4.36 &  0.436 &  6.04 &    2.69 &     1.42 &  9.43 &   5846  &  4.38  &  1.24 &  $+0.04$ & \\
090004 &  $-$5.25 &  0.176 &  1.76 &($<$1.2) &     1.33 &  9.83 &   5607  &  4.42  &  0.85 &  $-0.02$ & PHS \\
091287 &  $-$5.07 &  0.185 &  2.05 &    1.74 &     1.34 &  9.21 &   5648  &  4.46  &  0.88 &  $-0.01$ & \\
096184 &  $-$5.34 &  0.165 &  2.03 &    1.65 &     1.40 &  9.25 &   5863  &  4.45  &  1.00 &  $+0.13$ & \\
096395 &  $-$5.02 &  0.218 &  2.30 &    2.18 &     1.25 &  9.76 &   5816  &  4.48  &  1.00 &  $-0.10$ & \\
096402 &  $-$5.32 &  0.182 &  1.99 &($<$1.1) &     1.28 &  9.90 &   5661  &  4.20  &  1.00 &  $-0.03$ & \\
096901 &  $-$5.38 &  0.185 &  1.97 &($<$1.1) &     1.37 &  9.77 &   5742  &  4.32  &  1.01 &  $+0.08$ & PHS \\
096948 &  $-$5.21 &  0.175 &  1.85 &    1.52 &     1.30 &  9.62 &   5725  &  4.36  &  1.07 &  $+0.07$ & \\
097420 &  $-$4.72 &  0.209 &  2.57 &    2.22 &     1.41 &  9.65 &   5780  &  4.42  &  1.04 &  $+0.05$ & \\
098921 &  $-$4.38 &  0.329 &  3.66 &    2.47 &     1.39 &  9.13 &   5810  &  4.50  &  1.19 &  $+0.17$ & \\
100963 &  $-$4.97 &  0.200 &  2.39 &    1.72 &     1.24 &  9.71 &   5779  &  4.46  &  0.98 &  $ 0.00$ & \\
104075 &  $-$4.31 &  0.409 &  5.01 &    2.72 &     1.32 &  9.20 &   5881  &  4.37  &  1.08 &  $+0.05$ & \\
109110 &  $-$4.35 &  0.409 &  4.78 &    2.49 &     1.41 &  9.28 &   5835  &  4.51  &  1.11 &  $+0.07$ & \\
110205 &  $-$5.31 &  0.216 &  1.98 &    1.09 &     1.12 &  9.96 &   5708  &  4.28  &  1.08 &  $-0.23$ & \\
112504 &  $-$5.06 &  0.209 &  1.99 &    2.01 &     1.37 &  9.63 &   5741  &  4.34  &  1.00 &  $+0.01$ & \\
113579 &  $-$4.20 &  0.573 & 10.31 &    3.06 &     1.49 &  9.57 &   5759  &  4.21  &  1.44 &  $+0.05$ & \\
113989 &  $-$5.17 &  0.222 &  1.94 &($<$0.8) &     0.99 & 10.13 &   5506  &  4.38  &  0.74 &  $-0.46$ & outlier in [Fe/H]/$T_{\rm eff}$\\
115715 &  $-$5.25 &  0.215 &  2.34 &    1.72 &     1.22 & 10.08 &   5684  &  4.15  &  1.05 &  $-0.19$ & \\
116613 &  $-$4.32 &  0.349 &  2.96 &    2.12 &     1.22 &  8.95 &   5869  &  4.49  &  1.11 &  $+0.16$ & \\
Sun/Vesta&$-$5.33 &  0.193 &  2.29 &    0.92 &     1.22 &  9.66 &   5761  &  4.44  &  1.00 &  $-0.01$ & \\
\end{longtable}


\begin{thebibliography}{}
\bibitem[]{}
  Baliunas, S. L., et al. 1995, ApJ, 438, 269
\bibitem[]{}
  Baliunas, S., \& Jastrow, R. 1990, Nature, 348, 520
\bibitem[]{}
  Cox, A. N. 2000, Allen's Astrophysical Quantities, 4th ed. (Berlin: Springer)
\bibitem[]{}
  Duncan, D. K., et al. 1991, ApJS, 76, 383
\bibitem[]{}
  Eddy, J. A. 1976, Science, 192, 1189
\bibitem[]{}
  Hall, J. C., \& Lockwood, G. W. 2004, ApJ, 614, 942
\bibitem[]{}
  Hall, J. C., Lockwood, G. W., \& Skiff, B. A. 2007, AJ, 133, 862
\bibitem[]{}
  Isaacson, H., \& Fischer, D. 2010, ApJ, 725, 875
\bibitem[]{}
  Kurucz, R. L. 1993, Kurucz CD-ROM, No. 13 
  (Harvard-Smithsonian Center for Astrophysics) 
  [also available at 
  \texttt{http://kurucz.harvard.edu/PROGRAMS.html}]
\bibitem[]{}
  Kurucz, R. L., \& Bell, B. 1995, Kurucz CD-ROM, No. 23 
  (Harvard-Smithsonian Center for Astrophysics) 
  [also available at 
   \texttt{http://kurucz.harvard.edu/LINELISTS.html}]
\bibitem[]{}
  Noguchi, K., et al. 2002, PASJ, 54, 855
\bibitem[]{}
  Noyes, R. W., Hartmann, L. W., Baliunas, S. L., Duncan, D. K.,
  \& Vaughan, A. H. 1984, ApJ, 279, 763
\bibitem[]{}
  Strassmeier, K., Washuettl, A., Granzer, Th., Scheck, M.,
  \& Weber, M. 2000, A\&AS, 142, 275
\bibitem[]{}
  Takeda, Y., Honda, S., Kawanomoto, S., Ando, H., \& Sakurai, T. 2010, 
  A\&A, 515, A93 (Paper II)
\bibitem[]{}
  Takeda, Y., Kawanomoto, S., Honda, S., Ando, H., \& Sakurai, T. 2007, 
  A\&A, 468, 663 (Paper I)
\bibitem[]{}
  Takeda, Y., Tajitsu, A., Honda, S., Kawanomoto, S., Ando, H., 
  \& Sakurai, T. 2011, PASJ, 63, 697 (Paper III)
\bibitem[]{}
  Vaughan, A. H., \& Preston, G. W. 1980, PASP, 92, 385
\bibitem[]{}
  Vaughan, A. H., Preston, G. W., Wilson, O. C. 1978, PASP, 90, 267
\bibitem[]{}
  Wright, J. T. 2004, AJ, 128, 1273
\bibitem[]{}
  Wright, J. T., Marcy, G. W., Butler, R. P., \& Vogt, S. S. 2004,
  ApJS, 152, 261
\end{thebibliography}
\end{document}